\documentstyle[12pt]{article}

\newcommand{\Section}[1]{\section{#1}\setcounter{equation}{0}}

         \def\be{\begin{equation}}
         \def\bea{\begin{eqnarray}}
         \def\ee{\end{equation}}
         \def\eea{\end{eqnarray}}
         \def\ov{\over}
         \def\R{\rm {I\kern-.200em R}}
         \def\C{\rm {I\kern-.520em C}}

\begin{document}
\begin{tabbing}
  \` IPM-94-054
  \\
  \
  \\
\end{tabbing}
\vskip 10mm
{\large {\bf
          \centerline{
                    Baryogenesis from Long Cosmic Strings
                        } 
         } 
 } 
\vskip 10 mm

\centerline{
              Masoud Mohazzab\footnote{
                        e-mail addresses
                        masoud@netware2.ipm.ac.ir and masoud@irearn.bitnet}
             }
\vskip 10 mm
{\it
  \centerline{
              Institute for Studies in Theoretical Physics and
              Mathematics,
             }
  \centerline{
             P.O.Box  5746, Tehran 19395, Iran.
             }
    } 
\vskip 10 mm

\begin{abstract}

     Based on the mechanism of cusp production on long cosmic strings,
     the baryon asymmetry caused by cusp annihilation has been
     calculated. The result is compatible with observation and stronger than
     the results from loops.

\end{abstract}

\newpage

\Section{
          Introduction
          }

  Grand Unified Theories (GUT) are possible frameworks to explain the
  baryon asymmetry of the universe where the Sakharov's conditions \cite{S}
  i.e. 1) baryon number violation, 2) C and CP violation and 3) departure
  from thermal equilibrium are satisfied. GUTs predict that the universe
underwent
  a series of phase transitions during its early stage of evolution.
  Cosmic strings are one dimensional topological defects that are
  generated during the phase transition \cite{K}, \cite{VIL}.
  Cosmic strings have to be in the
  form of loops or infinitely long with the important parameter $\mu $,
   mass per length of the strings and are very desirable in cosmology.
   The value $G \mu \sim 10^{-6}$ is compatible with \\
  1) the scale of GUT symmetry breaking \\
  2) structure formation models and     \\
  3) thermal fluctuations of the background radiation (the recent COBE
results). \

  The oscillations of cosmic strings typically lead to the formation of cusps.
  A cusp is a
  point where two segments of the strings overlap and the point reaches the
  speed of light. There is no topological barrier to conserve cusps from
  decaying into bursts of Higgs or superheavy particles. The rate
  of decay and total energy of cusp annihilation has been worked out in
  \cite{ST} and \cite{M}, respectively. The starting point for the perturbative
  calculation is to consider the interacting lagrangian as [3]
  \be
  {\cal L}_I = \lambda \vert \phi \vert^2 \psi^2
  \ee
  where $\lambda $ is the coupling constant of the Mexican hat potential
  \be
  U(\phi ) ={1\ov 4} \lambda (\vert \phi \vert^2 -\sigma^2)^2
  \ee
   $\psi $ is the outgoing particles, $\sigma $ is the scale of GUT symmetry
breaking
   and $\phi $ is the complex field configuration of the string.

The dimensionless value that most authors try to derive is the ratio of the net
baryon number density, denoted by $n_B - n_{\bar B} $, to the entropy
density of the universe, $s $, or $\eta ={{n_B - n_{\bar B}}\ov s} $. The
observational value for $\eta $ based on the measurments of aboundances of
primordial
D, $^3$He and $^7$Li is in the range $(6 - 10) \times 10^{-11} $.

Cusp annihilation on cosmic strings has long been considered as a
mechanism to describe the baryon asymmetry of the universe \cite{KM}.
 The annihilation of cusps produce superheavy $\psi$ particles. Furthere  decay
of
 these particles
results in the baryon asymmetry of $\Delta  B \sim 10^{-2} - 10^{-13}$
\cite{NW}.
In \cite{KM} loops of ordinary cosmic strings are considered and the baryon
number
production from kinks and cusps evaporation has been calculated.
The result of \cite{KM} is independent of era shows cusp
annihilation on loops results in a compatible value for $\eta $.
Kink evaporation, on the other hand, cannot describe the observed
value of baryon asymmetry.
 Another mechanism to be accounted for
baryogenesis is the collapse of the topological defects such as cosmic
string loops or textures \cite{BDH}. In the former approach, oscillations of
cosmic string loops reduce their radius due to gravitational radiation.
The loops shrink to their minimum size $\sim \lambda^{-{1\ov 2}} \sigma^{-1} $
when they have zero winding number and can decay into superheavy particles.
The decay of these superheavy particles are accounted for the baryon number
asymmetry.  The other mechanism of \cite{BDH} is the shrinking of the loops by
cusp annihilation. In each period of oscillation of a loop, equivalent to its
length,
a cusp forms and radiates causing the loss of energy and therefore the
decrease of the radius of the loop.

The ratios of baryon number generation of loops due to the decrease of its
radius by gravitational radiation and cusp annihilation are respectively

\be  \label{b1}
\eta_{g.r.}=\eta_{max} (G\mu )^{-{3\ov 4}} ({T_F\ov T_c})^3
\ee

\be \label{b2}
\eta_{c.a.} =\eta_{max} \lambda^{1\ov 2} ({T_F\ov T_c})^{3/2}
\ee
where $\eta_{max} \simeq .03 {N_x\over N} \Delta B$ ( $10^{-13} < \Delta B
<10^{-2}$)
and $N_x$ is the helicity of spin and N is the number of states.
${T_F}$ is the temperature at the time $t_F =(\gamma  G \mu)^{-1} t'_F$
where $t_F$ is the time that loops reaches to its minimum size length and
decays
and $t'_F$ is the time when the loop is formed.

 In \cite{MB} a mechanism has been suggested that produces cusps from colliding
 traveling waves on long strings. In this work, based on the same mechanism, we
 calculate the baryon asymmetry caused by long cosmic strings from cusp
 annihilation.

 In the next section, we calculate the
 amount of baryon number from cusp annihilation on long strings.

\section {
 Baryon number production from long strings }

According to  a numerical simulation, long strings are ${80\%} $ of the
total cosmic strings formed at the GUT
phase transition \cite{VIL}. Therefore they may have important contribution
to radiation from cosmic strings.

Long strings do not shrink to a point and the only
possibility for them to radiate superheavy particles is through cusp
annihilation.
 Cusps, on long strings, are formed, up to a probability, when two wiggles
traveling along the string collide.
If we assume the wiggles have
random shapes we can find the probability of formation of a cusp is $50\%$
\cite{MB}.
 The probability of formation of more than one cusp in each collision is also
non-zero. Of course it
is clear that the shapes of wiggles are not random since the fractal dimension
of cosmic strings is $1.2$ (and not 2 for the random shape)\cite{BBT}. Due to
this mechanism and the possibility of superheavy particle production from cusps
annihilation, long strings could contribute to the baryon number generation.

The scaling distribution of long strings is given by \cite{VIL}

\be  \label{sc}
n_{l.s.}^{scaling}={\nu\over t^3}
\ee
 where $\nu $ is a constant of the order 100 and $t$ is the cosmological time.
 The distribution before the scaling solution, however, is different and is
 given by \cite{KE}

 \be   \label{ns}
n_{l.s.}={\nu' \ov {{(G\mu )^{3\ov 2} m_{pl}^{3\ov 4} t^{3+{3\ov 4}}}}}
\ee
where $\nu' $ is constant.

The distribution of traveling waves on long strings can roughly be written
\cite{MB} as

\be
K(l,t)=\alpha {t\over l^2}
\ee
where $l$ is the size (the order of wavelength) of the traveling wiggle
and $\alpha $ is a constant.

The total number of cusps per volume will be

\be  \label {bb}
n_c(t)=P_c \int^{t}_{t_{min}} dt' n_{l.s.} \int^{t'}_{l_{min}} dl K(l,t') {1\ov
l} z(t')^{-3}
\ee
where $z(t)$ is the red shift factor at the time $t $,  ${1\ov l}$
represents the frequency of the impacts of the wiggles.
The probability of cusp formation on
long string upon each collision is denoted by ${P_c}$.

In the integration (\ref {bb}) $l_{min}$ is a minimum for the size
of traveling waves, beyond which gravitational radiation smoothes out
the wiggle

\be
    l_{min} \simeq \gamma G \mu t
 \ee
where $\gamma \simeq 100$.

The minimum time $t_{min}$ is close to the time when cosmic string is formed,
i.e. $t_c $,
the time when GUT phase transition occurs.

It should be noted that we have neglected the effect of friction and the fact
that the traveling waves do not change their shape when they travel along the
string \cite{VV}.

  To implement the third Sakharov condition, i.e. out of equilibrium
  condition, we should have

\be  \label{lt}
\Gamma \le H
\ee
where $\Gamma $ is the decay rate of the superheavy particle and $H$ is
the expansion rate of the Universe. Here we consider the decay
of superheavy particles and don't consider the inverse decay or collisions.
Therefore the minimum time can be approximated by $t_{min}= t_s \sim 10
m_{\psi}^{-1} $ \cite{KM},
when the inverse decay is suppressed by boltzman factor. $t_s $ is given by
\be \label{ts}
t_s= 100 \sqrt{{45\ov {16\pi^3 g_* }}} ({m_{\psi } \ov m_{pl}})^{-1} t_{pl}
\ee

Using the number density (\ref{ns}), the total number of cusps at the time $t$,
 before the scaling solution era, will be

$$n_c(t)={7\nu' P_c \alpha \over 2(\gamma G \mu)^2(G\mu)^{3\ov2}m_{pl}^{3\ov
4}} {1\ov t^2} {1\over t_s^{7\ov 4}}$$

 The net baryon number from cusp evaporation is

\be
n_B= n_c \Delta B
\ee
where $\Delta B $ is the amount of baryon number violation due to CP violation
from the decay of the superheavy particles and is in the range of
$10^{-2} - 10^{-13}$, depending on the model \cite{NW}.

 The baryon number asymmetry will then be

\be \label{et}
\eta \equiv {n_B \over {s}} ={7\nu' P_c \alpha \Delta B\over 2(\gamma G
\mu)^2(G\mu)^{3\ov 2}}{1\ov (.30118 g_*^{-{1\ov 2}})^2}{1\ov {2\ov 45}\pi^2
g_{*s}}{T\ov m_{pl}} {1\over (m_{pl}t_s)^{7\ov 4}}
 \ee
$s$ is the entropy density
$s={ 2 \ov 45}  \pi^2 g_{*s} T^3$, $ g_{*s}$ is the number of
states and $T $ is the temperature of the universe at the time $t$.

 By (\ref{et}) it is seen that the baryon asymmetry produced
by long strings depends on the era. It is easy to see, however,
that the asymmetry produced in the early era is much more important. In fact
by using (\ref{sc}) instead of (\ref{ns}), in the integration (\ref{bb}), and
performing the integration up to the present time it can be seen that the
value of asymmetry is $30$ order of magnitude less than the observed value.
As a result we just consider the pre-scaling era. Another result of (\ref{et}
is that the suppression factor ( i.e. the exponent of $T$) is $1$ while in the
case of loops it is more severe
 (equations (\ref{b1}) and (\ref{b2})\cite{BDH} ).
For $\nu' \sim 1$, $\gamma G \mu = 10^{-4}$, $G \mu = 10^{-6}$, $g_*=
g_{*s}=100$
and using (\ref{ts}) we will have

\be
\eta= .566 \times 10^{11}\alpha P_c \Delta B {T\ov m_{pl}} ({m_{\psi } \ov
m_{pl}})^{7\ov 4}
\ee

The exact value for $P_c$ and $\alpha $ is not known yet, but it is surely
$P_c< {1\ov 2}$. In \cite{MB} it was assumed $\alpha = 1$ which is an upper
estimate for $\alpha $ and we should have $\alpha < 1 $.

Therefore, for $m_{\psi } \sim 10^{15} GeV $ \cite{M}, $ \alpha P_c = .5$
\cite{MB},
$T \sim 10^{13}$ GeV and  $m_{pl} \sim 1.2 \times 10^{19}$ GeV  we see
that for the value $\Delta B \sim 10^{-8}$, the baryogenesis from long cosmic
strings is consistent with the observed value.

\section {
           Conclusion }

Long strings are able to radiate superheavy particles via production and
annihilation
of cusps. Cusps may form on long strings when two traveling wiggles collide.
The annihilation of cusps could be accounted for baryon asymmetry of the
universe. By calculating the total number of cusps on long strings, as a
function of time, we have calculated the amount of baryon asymmetry from them.
It is assumed that one (or a pair of) superheavy gauge particle(s) is produced
from each cusp annihilation.
The suppression factor for long strings is less than that of loops. As a result
long strings have more important contribution than loops to the baryon
asymmetry.

\section*{Acknowledgments}

I am grateful to Farhad Ardalan for useful comments.


\begin{thebibliography}{99}
\bibitem{S}
A. D. Sakharov, ZhETF Pis'ma {\bf 5} (1967) 32 [JETP Lett. {\bf 5} (1967) 24.]
\bibitem{K}
 T. W. B. Kibble Phys. Rep. 67 (1980) 183.
\bibitem{VIL}
A. Vilenkin
Phys. Rep. 121 (1985) 263.
\bibitem{ST}
M. Serendicki and S. Theisen, Phys. Lett. {\bf B 189} (1987) 397; R. H.
Brandenberger
, Nucl. Phys. {\bf B 293} (1987) 812.
\bibitem{M}
M. Mohazzab, Int. J. Mod. Phys. {\bf D} (1993).
\bibitem{KM}
M. Kawasaki M. and K. Maeda
Phys. Lett. {\bf B 209} (1988) 271.
\bibitem{NW}
D. Nanopoulos and S. Weinberg, Phys. Rev. {\bf D 20} (1979) 2484;
A. Yildiz and P. H. Cox, Phys. Rev. {\bf D 21} (1980) 906;
P. Langacker in "CP violation" edited by C. Jorlskog, World Scientific, 1989.
\bibitem{BDH}
R. H. Brandenberger, A. C. Davis and M. Hindemarsh
Phys. Lett. {\bf B 263} (1991) 239.
\bibitem{MB}
M. Mohazzab and R.H. Brandenberger
Int. J. Mod. Phys. {\bf D 2} (1993) 185.
\bibitem{BBT}
F. R. Bouchet, "The formation and evolution of cosmic strings", editted by
G. Gibbons, S. Hawking and T. Vachespati, Cambridge University Press 1990;
D. P. Bennett, {\it ibid}; N. Turok, {\it ibid}.
\bibitem{KE}
A. Everett, Phys. Rev.{\bf D 24} (1981) 858;
T. W. B. Kibble, Acta Phys. Pol. B 13 (1982) 723.
\bibitem{VV}
Vachespati and T. Vachespati
Phys. Lett. {\bf B 238} (1990) 41;
Grafinkle and T. Vachespati
Phys. Rev.  {\bf D 42}, (1990) 1960.

\end{thebibliography}
\end {document}